\documentstyle[epsf,epsfig,aps]{revtex}
\renewcommand{\thefootnote}{\fnsymbol{footnote}}
\begin{document}
\newcommand{\be}{\begin{eqnarray}}
\newcommand{\dlq}{\lq\lq}
\newcommand{\ee}{\end{eqnarray}}
\newcommand{\ben}{\begin{eqnarray*}}
\newcommand{\een}{\end{eqnarray*}}
\newcommand{\beq}{\begin{equation}}
\newcommand{\eeq}{\end{equation}}
\renewcommand{\baselinestretch}{1.0}
\newcommand{\as}{\alpha_S}
\def\eq#1{{Eq.~(\ref{#1})}}

\def\ap#1#2#3{     {\it Ann. Phys. (NY) }{\bf #1} (19#2) #3}
\def\arnps#1#2#3{  {\it Ann. Rev. Nucl. Part. Sci. }{\bf #1} (19#2) #3}
\def\npb#1#2#3{    {\it Nucl. Phys. }{\bf B#1} (19#2) #3}
\def\plb#1#2#3{    {\it Phys. Lett. }{\bf B#1} (19#2) #3}
\def\prd#1#2#3{    {\it Phys. Rev. }{\bf D#1} (19#2) #3}
\def\prep#1#2#3{   {\it Phys. Rep. }{\bf #1} (19#2) #3}
\def\prl#1#2#3{    {\it Phys. Rev. Lett. }{\bf #1} (19#2) #3}   
\def\ptp#1#2#3{    {\it Prog. Theor. Phys. }{\bf #1} (19#2) #3}
\def\rmp#1#2#3{    {\it Rev. Mod. Phys. }{\bf #1} (19#2) #3}
\def\zpc#1#2#3{    {\it Z. Phys. }{\bf C#1} (19#2) #3}
\def\mpla#1#2#3{   {\it Mod. Phys. Lett. }{\bf A#1} (19#2) #3}
\def\nc#1#2#3{     {\it Nuovo Cim. }{\bf #1} (19#2) #3}
\def\yf#1#2#3{     {\it Yad. Fiz. }{\bf #1} (19#2) #3}
\def\sjnp#1#2#3{   {\it Sov. J. Nucl. Phys. }{\bf #1} (19#2) #3}
\def\jetp#1#2#3{   {\it Sov. Phys. }{JETP }{\bf #1} (19#2) #3}
\def\jetpl#1#2#3{  {\it JETP Lett. }{\bf #1} (19#2) #3}
\def\epj#1#2#3{    {\it Eur. Phys. J. }{\bf C#1} (19#2) #3}
\def\ijmpa#1#2#3{  {\it Int. J. of Mod. Phys.}{\bf A#1} (19#2) #3}
\def\ppsjnp#1#2#3{ {\it (Sov. J. Nucl. Phys. }{\bf #1} (19#2) #3}
\def\ppjetp#1#2#3{ {\it (Sov. Phys. JETP }{\bf #1} (19#2) #3}
\def\ppjetpl#1#2#3{{\it (JETP Lett. }{\bf #1} (19#2) #3}
\def\zetf#1#2#3{   {\it Zh. ETF }{\bf #1}(19#2) #3}
\def\cmp#1#2#3{    {\it Comm. Math. Phys. }{\bf #1} (19#2) #3}
\def\cpc#1#2#3{    {\it Comp. Phys. Commun. }{\bf #1} (19#2) #3}
\def\dis#1#2{      {\it Dissertation, }{\sf #1 } 19#2}
\def\dip#1#2#3{    {\it Diplomarbeit, }{\sf #1 #2} 19#3 }
\def\ib#1#2#3{     {\it ibid. }{\bf #1} (19#2) #3}
\def\jpg#1#2#3{        {\it J. Phys}. {\bf G#1}#2#3}

\begin{flushright}
TAUP  2651-2000\\
\today \\
\end{flushright}
\vspace*{1cm}
\setcounter{footnote}{1}
\begin{center}
{\Large\bf Low $\mathbf{x}$ Physics}
\\[1cm]
 Eugene
Levin  \\ ~~ \\

{\it $^3$ HEP Department, School of Physics and Astronomy } \\
{\it Tel Aviv University, Tel Aviv 69978, Israel } \\
{\it and}\\
{\it DESY Theory,  22603 Hamburg, Germany}
~~ \\ ~~ \\
\centerline{\it Talk, given at Diffraction'2000, Sept. 2 - 7, Centraro,
Italy}

~~\\
~~\\

\end{center}
\begin{abstract}
In this talk, we present the arguments, that a new QCD regime - gluon
saturation, has been  reached at HERA.
\end{abstract}

\renewcommand{\thefootnote}{\arabic{footnote}}
\setcounter{footnote}{0}

\section{Main questions and problems}

~
We hope that everybody agrees,  we have two principle problems in low
$x$ physics: (i)  matching  between ``soft" (npQCD)  and ``hard"(pQCD)
processes; and (ii) theoretical description of high parton  density QCD
(hdQCD).
My credo is \cite{GLR,MUQI,LV}: 
{  \em    
these two problems are correlated
and the system of partons always passes the stage of hdQCD
( at shorter distances ) before it goes to the black box, which we call
non-perturbative QCD and which, in  practice, we describe in old fashion
Reggeon phenomenology.}

In this talk you can find the answer to the following questions:
\begin{itemize}
\item \quad Is there any violation of the DGLAP evolution?
\item \quad Can we describe  matching between ``soft" and ``hard"
processes?
\item \quad  Do we see a signal of  the  gluon saturation in HERA data?
\item \quad What is a current situation in theory for high parton density
QCD system?
\end{itemize}

Unfortunately, the lack of space does not allow us to discuss such a hot
problem as the status of the BFKL Pomeron as well as the manifestation of
high parton density QCD in the Tevatron data.

\section{ Two scales of DIS}

About thirty years ago Gribov\cite{GRIB} noticed that photon - hadron
interaction   at high energies has two distinct stages as far as time -
space picture of interaction is concerned:

\begin{figure}
 \epsfig{file=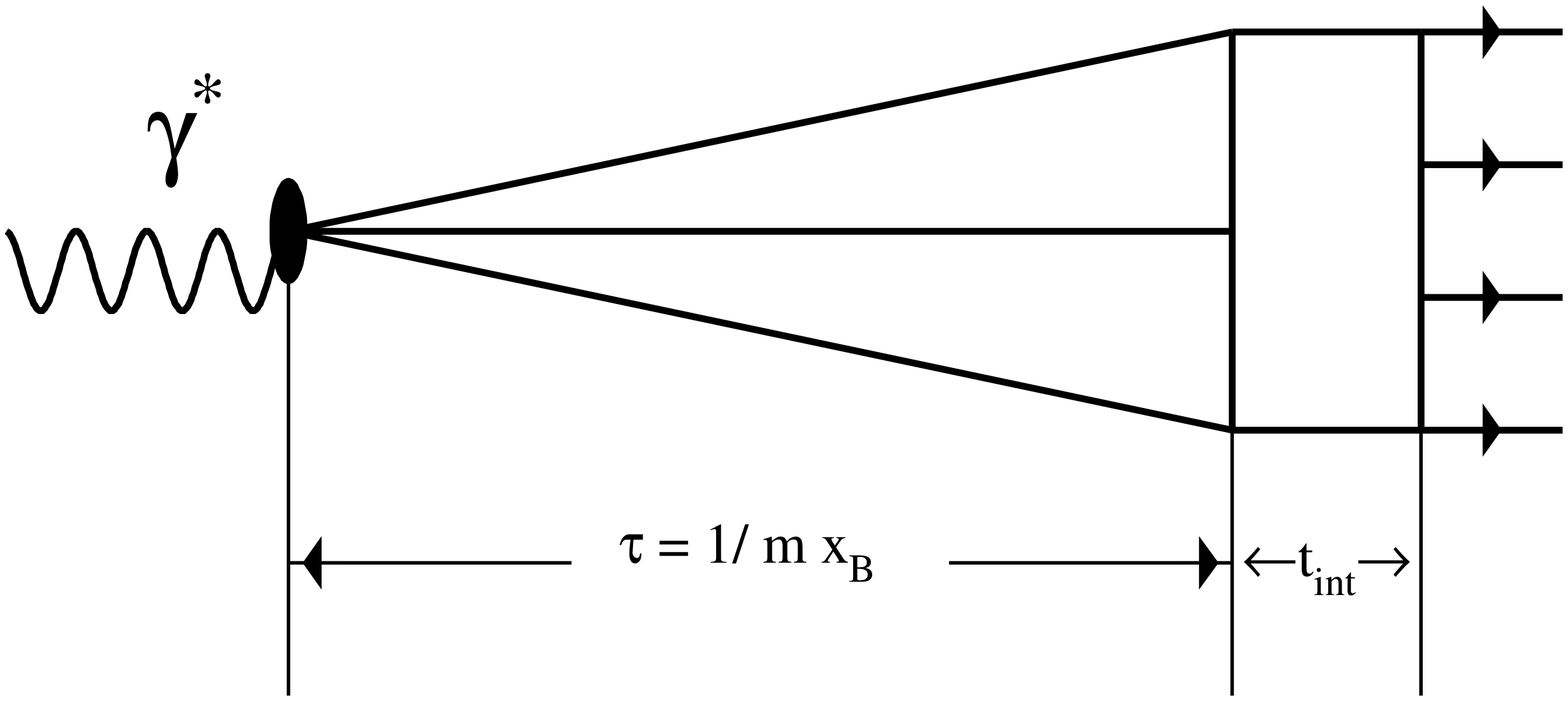, width=12cm}
\label{time}
\caption{}
  \end{figure}

\begin{enumerate}
\item\,\,\,$\gamma^*  \,\,\longrightarrow$
  hadron system ( $q \bar q $ - pair );
\item\,\,\,  hadron system ( $q \bar q $ - pair )
interacts with the target.
\end{enumerate}
Therefore, we can describe photon-hadron interaction as follows
\beq \label{GENFOR}
\sigma_{tot} ( \gamma^* p )\,\,=\,\,\sum_{n} \,|\Psi_n|^2 \,\,\sigma_{tot}
( n p)\,\,,
\eeq
where $n$ denotes the set of quantum numbers which diagonalize the high
energy interaction matrix. This set of quantum numbers we call the
correct degrees of freedom (DOF) and $ \sigma_{tot}(n p)$ is the total
cross section for the interaction of the hadron or parton system with
quantum numbers $n$ with the target. \eq{GENFOR} is useful only if we know
the correct DOF. Fortunately, we do know them at short distances where $n$
are   colour dipoles \cite{MU94} and \eq{GENFOR} reads as 
 \beq \label{CD}
\sigma_{tot} ( \gamma^* p ) =  \int d^2 r_t
\int^1_0 \,d z\, | \Psi( Q^2; r_t,z ) |^2 \,\sigma_{tot} ( r^2_t, x
)\,\,.
\eeq
\subsection{ Separation scale $\mathbf{r^{sep}_{\perp}= 1/M_0}$}

However, at long distances we do not know the correct DOF. The scale which 
says what distances are short we call a separation scale (see Table.1).
\newpage
\centerline{\bf Table 1.}
\begin{center}
\begin{tabular}{ c c c}
{ $ r_{\perp} \,\,<\,\,$} & {  $ r^{sep}_{\perp}$}
&\,\,{ $<\,\,
r_{\perp}
$}\\
   & & \\
$\Psi( Q^2; r_t,z ) \,\,{ \rightarrow}$\,\,{ pQCD}& &
$\Psi( Q^2; r_t,z ) \,\,{\rightarrow}$\,\,{ npQCD}\\
& & \\
DOF: colour dipoles \cite{MU94} & & DOF: constituent quarks \cite{AQM}  \\
                      & & \,\,\,\,colour dipoles \cite{CDLONG} \\
& & \\
$\sigma_{tot} (r_{\perp},x) \propto $ & &
$\sigma_{tot} (r_{\perp},W) \rightarrow$  \\
 $r^2_{\perp}\,xG(x,4/r^2_{\perp})$ &  &Regge phenomenology\\

\end{tabular}
\end{center}

\subsection{ Saturation scale $\mathbf{r^{sat}_{\perp}
\approx\,1/Q_s(x)}$}

This scale can be estimated from the equation
\beq \label{KAPPA}
\kappa\,\,\,=\,\,\,\frac{3\,\pi^2 \alpha_S}{2
Q^2_s(x)}\,\times\,
\frac{xG(x,Q^2_s(x))}{\pi\, R^2}\,\,=\,\,1\,\,,
\eeq
which says that the packing factor of partons in the parton cascade is
about unity or, in other words, at this scale the parton system is
so dense that we cannot apply the standard methods of perturbative QCD.

The  physical meaning of $\kappa$ is clear from Fig. 2 which shows the
parton distributions in the transverse plane. At short distances $\kappa$
is the product of the parton density in transverse plane which is equal to
{\it number of partons/area} = $x G(x,Q^2)/\pi R^2$ multiplied by the
cross section of  the parton interaction $\propto \as/ Q^2$.
Therefore, at the scale where $\kappa \approx 1$ the interaction 
between partons become essential and new approach should be developed in
QCD to describe the high parton density system.
 
The natural hierarchy of the scales is $
r^{sat}_{\perp}\,\,{\ll}\,\,r^{sep}_{\perp}$ accordingly to
our main idea.
\section{ Theory Status}
The theory of high density system in QCD is in a very good shape now.
The problem has been attacked from two different point of view: (i) from 
the pQCD region\cite{GLR,MUQI,PQCD} by summing corrections to the
evolution equations due to
high  density of partons; and (ii) from the non-perturbative QCD region by
developing effective Lagrangian approach\cite{LV,NPQCD} dealing with such
a system. The resulting evolution equation that has been proven \cite{EQ}
looks as follows

\begin{eqnarray}
\frac{d a^{el}(\vec{x}_{01},b_t,y)}{d y}\,\,\,&=&\,\,\,- \,\frac{2  
\,C_F\,\as}{\pi} \,\ln\left( \frac{x^2_{01}}{\rho^2}\right)\,\,
a^{el}(\vec{x}_{01},b_t,y)\,\,\,
 +
\,\,\,\frac{C_F\,\as}{\pi}\,\,
\int_{\rho} \,\,d^2 x_{12}\,
\frac{x^2_{01}}{x^2_{02}\,
x^2_{12}}\,\nonumber \\
  &\cdot &\,\,\,
(\,\,2\,a^{el}(\vec{x}_{02}, \vec{ b}_t -
\frac{1}{2} \vec{x}_{12},y) \,\,\,
  - \,\,\,a^{el}(\vec{x}_{02},b_t,y)
\,\,a^{el}(\vec{x}_{12},b_t,y)\,\,)\,\,, \label{EQ} 
\end{eqnarray}
where $a^{el}$ is the elastic amplitude for dipole scattering at fixed
impact parameter $b_t$ and at energy $s$ ( $y = \ln s$).

\begin{figure}
\begin{center}
\epsfig{file=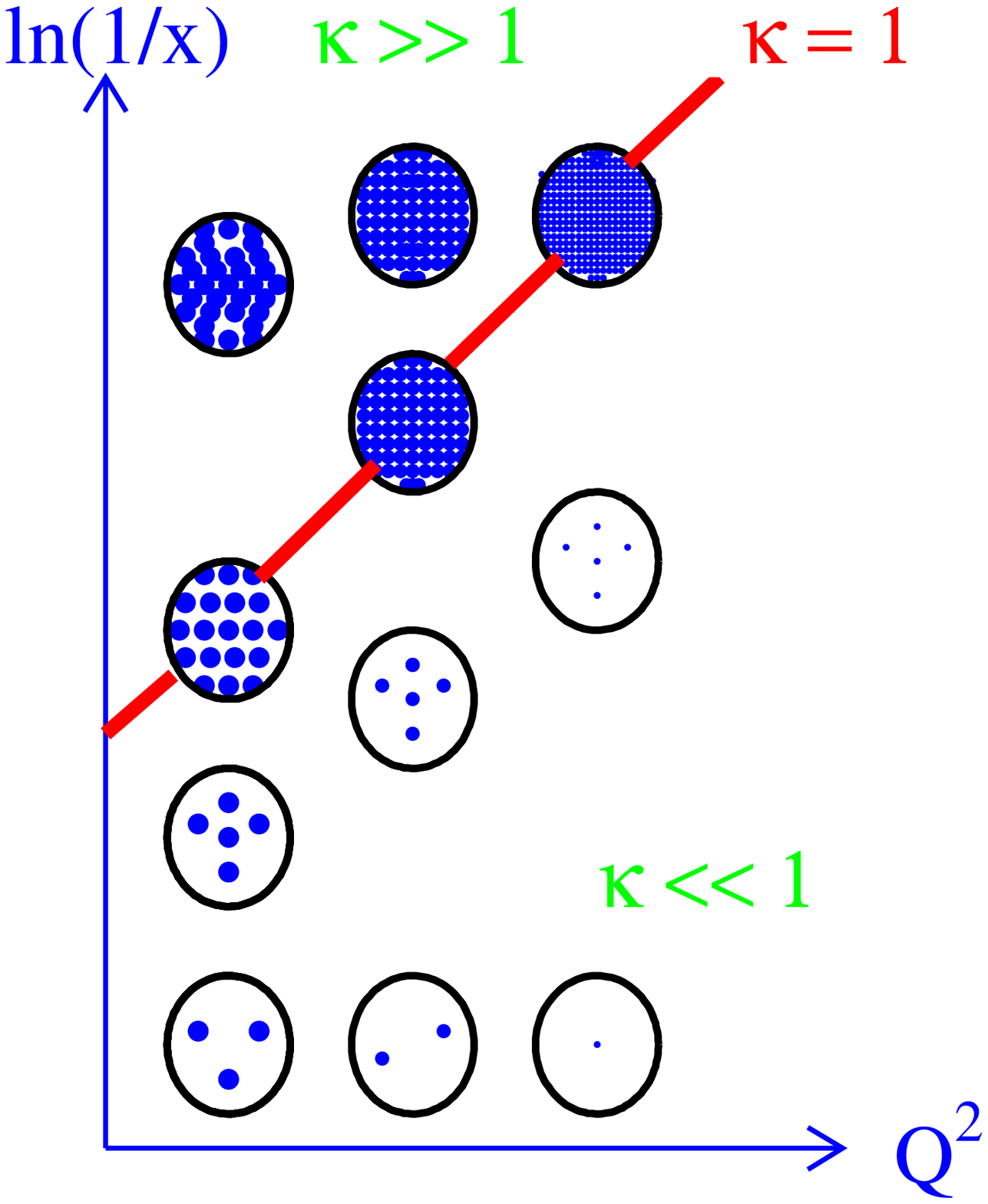,width=100mm}
\end{center}
\caption{}  
\end{figure}

\begin{figure}
\begin{center}
\epsfig{file=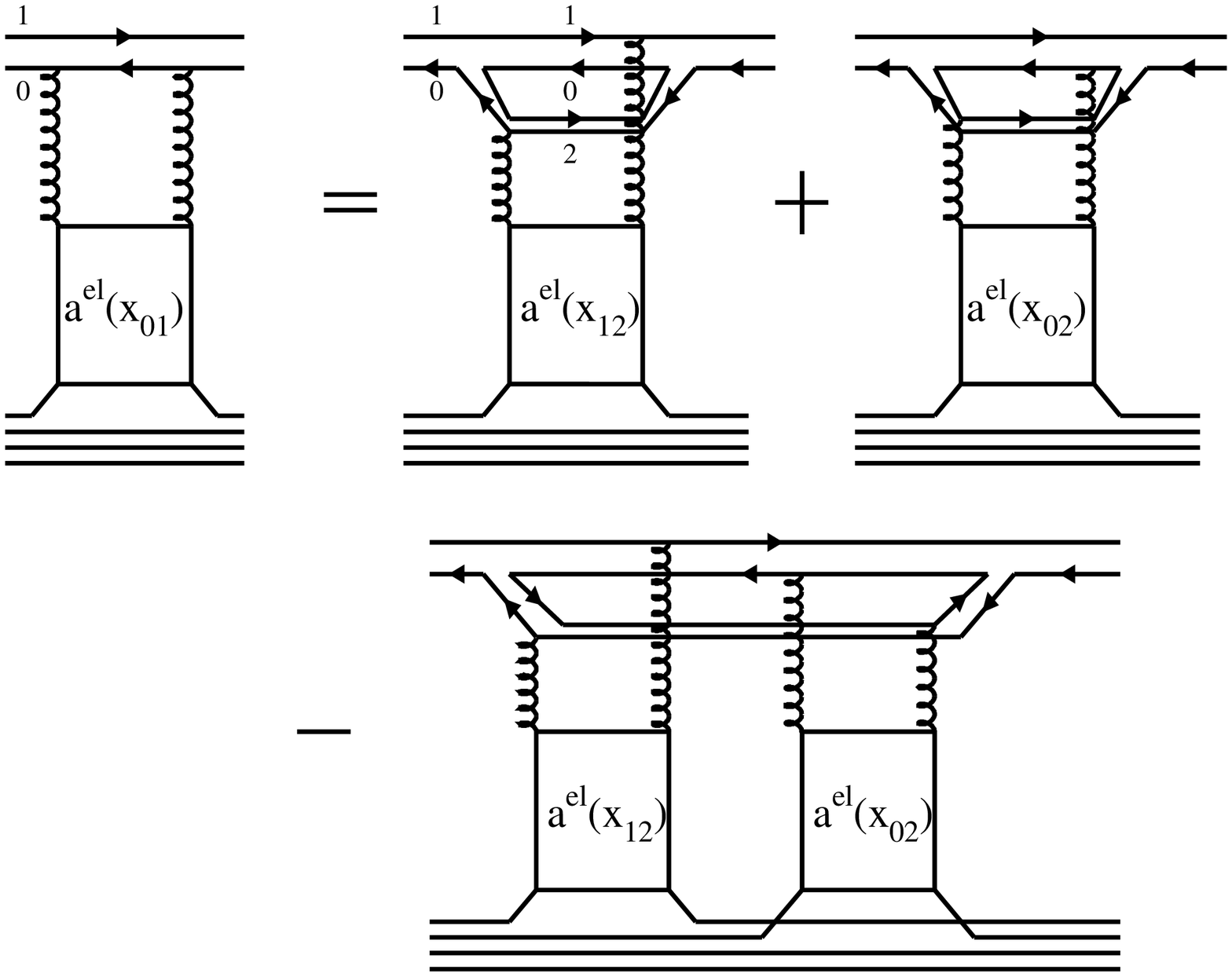,width=100mm}
\end{center}
\caption{}
\end{figure}

The physical meaning of \eq{EQ} is very clear from Fig. 3. Indeed, 
with probability $\frac{\mathbf{x^2_{01}}}{\mathbf{x^2_{02}}\,
\mathbf{x^2_{12}}}$ dipole with size $ \mathbf{x^2_{01}}$
 decays in two
dipoles with sizes $ \mathbf{x^2_{02}}$ and $ \mathbf{x^2_{12}}$ (see Fig.
3 ). These two dipoles interact with the target: each produced dipole
interacts with the target separately  ( the second term in \eq{EQ} ) or
two dipoles interact with the
target simultaneously ( the third term in  \eq{EQ} ). The first term
describes the fact that dipole $ \mathbf{x^2_{01}}$ disappears from the
initial state after  decay into two dipoles. This equation, which has been
suggested in the momentum representation in Ref.\cite{GLR}, 
  has
a lot of 
nice properties including the correct matching with the DGLAP evolution
equations. However, the most important message concerning this equation
is that this equation can be derived using the effective Lagrangian
approach \cite{LV}. It should be stressed that we know not only the
equation but also the initial condition for it.

\section{Matching of ``soft" and    ``hard" photon-proton interactions}

~ 
Serious attempts\cite{SEPSC} to find the value of the separation scale  $
r^{sep}_{\perp}\,\,=\,\,1/M_0$ were undertaken using Gribov's formula
\cite{GRIB} during the past decade, starting from pioneering paper of
Kwiecinski and Badelek \cite{KB}. Gribov's formula reads (see Fig. 5)
\beq \label{GRFOR}
\sigma(\gamma^* N)= 
\frac{\alpha_{em}}{3\pi}\int
  \frac{\Gamma(M^2) dM^2}{(Q^2+M^2)}
    \sigma(M^2,M'^2,s)
  \frac{\Gamma(M'^2) dM'^2}{(Q^2+M'^2)}
\eeq

\begin{figure}
\begin{center}
\epsfig{file= 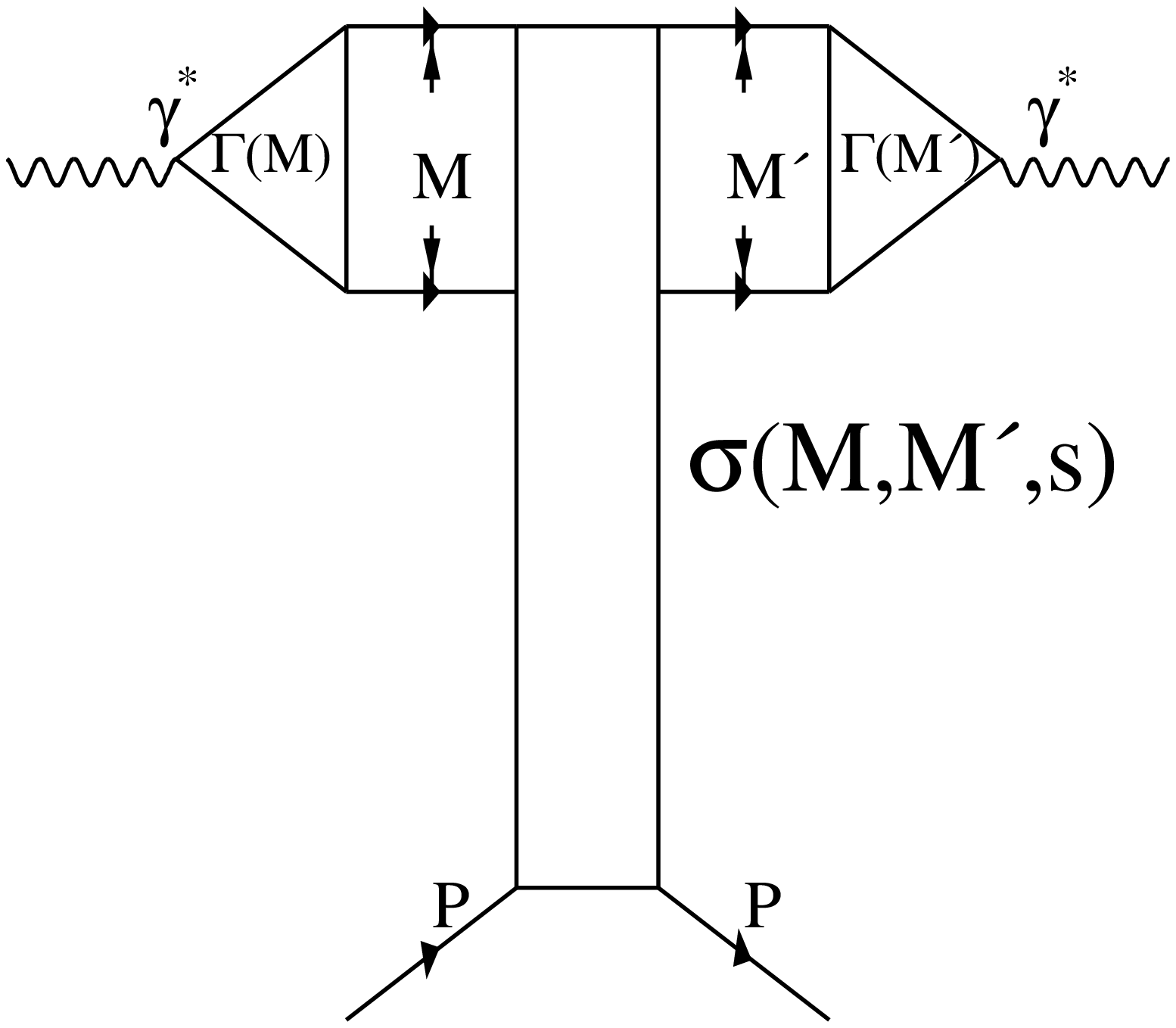, width= 10cm,height=4cm}
\end{center}     
 \label{grb}
\caption{}
  \end{figure}

~

The following assumptions are made to describe the matching between
``soft" and ``hard" processes:

~

~

\begin{tabular} {c c}
$ M < M_) $ &$ M > M_0$\\
 & \\
$M = M'$ & $M \neq  M'$ \\
 & \\
 D-L Pomeron + AQM\,\,\,\, & \,\,\,\,``Hard" Pomeron $\equiv$
pQCD approach\\
\end{tabular}

~

~

The result ( see Fig. 5 )   is that the separation scale depends on
polarization of
the incoming photon and it is equal
  $0.7\,\,<\,\,M^2_{0
T}\,\,<\,\,0.9\,\,GeV^2$ for transverse polarized photon and it is
smaller for the longitudinal polarized photon ( $M^2_{0
L}\,\,<\,\,0.4\,\,GeV^2$ ).

\section{ Where are SC?}

The HERA data put a puzzling question on the table:
\begin{itemize}
\item \quad On one hand, the data can be described by the routine DGLAP
evolution equations \cite{HERAREV};
\item \quad On the other hand, the gluon density measured at HERA is so
high that some effect of hdQCD should be seen (see Fig. 6 where parameter
$\kappa$ is plotted as it appears in HERA data. )

\begin{figure}
\begin{tabular}{l l}
\epsfig{file=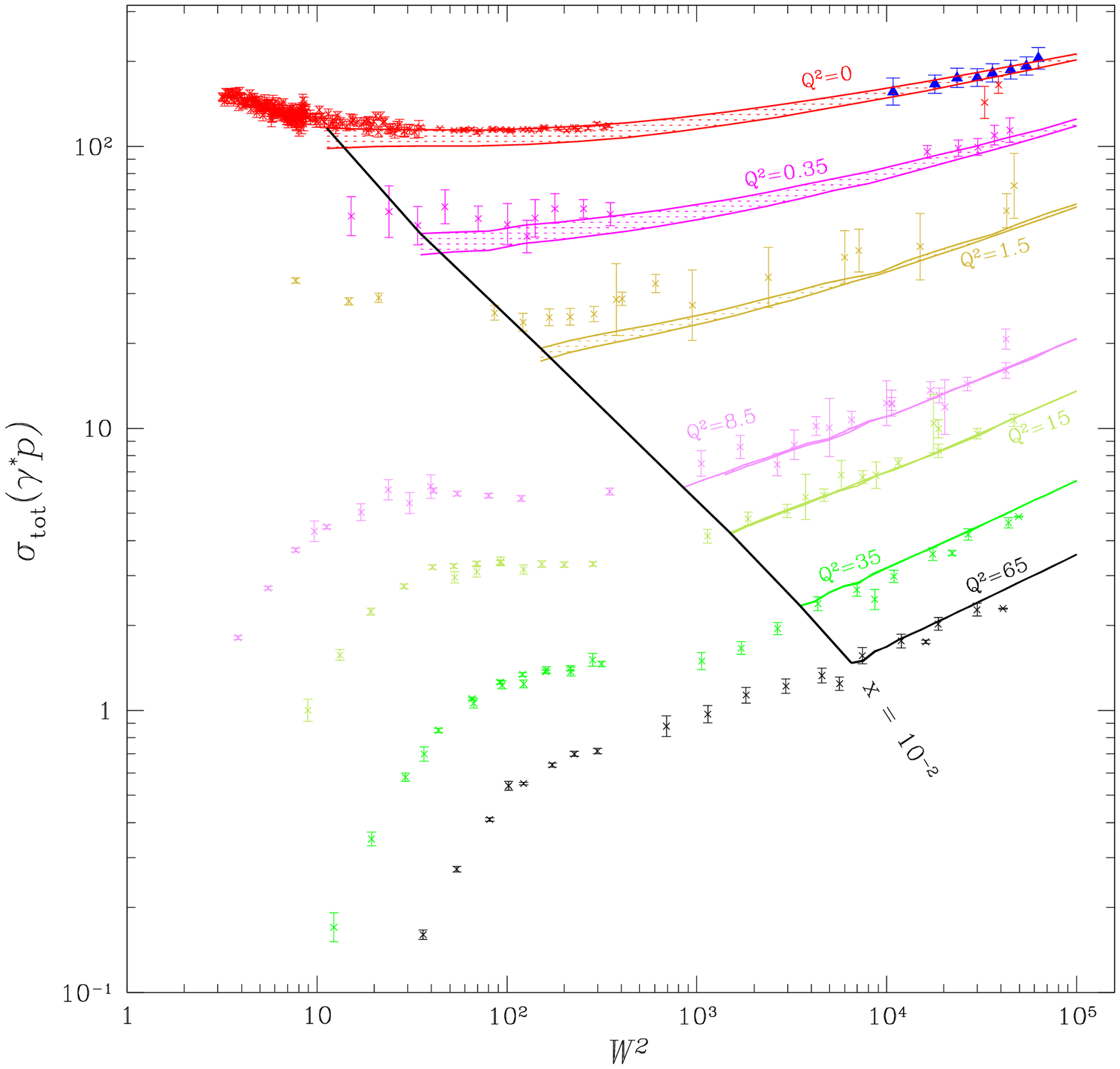,width=8.5cm} &
\epsfig{file=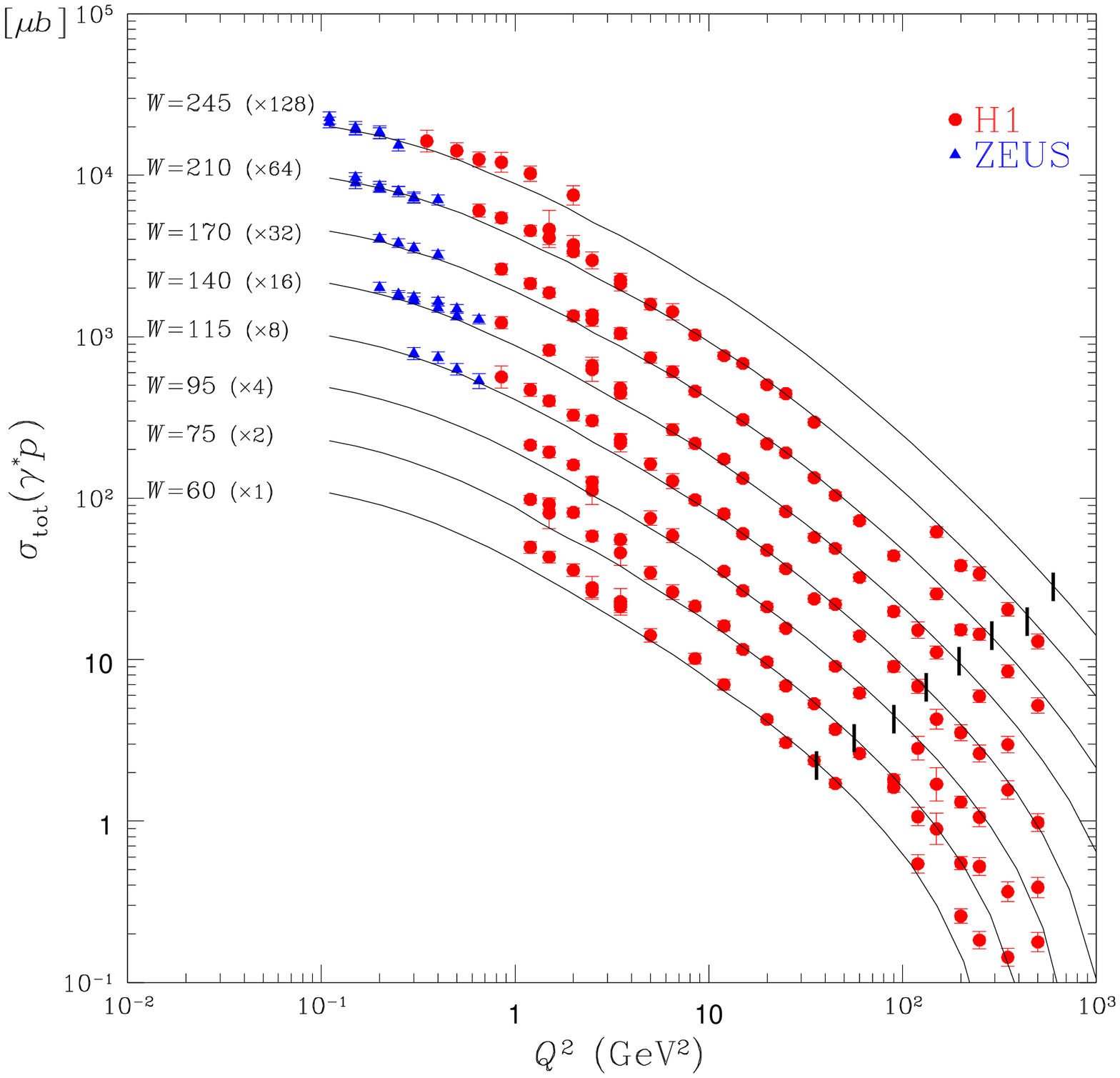,width=8.5cm}\\
\end{tabular}
\caption{ An example of description the HERA experimental data using
Gribov's formula, taken from GLMN paper \protect\cite{SEPSC}}
\end{figure}

The natural question to ask is where  to look for a saturation scale
$Q_s(x)$ . Fig.  7 shows us that the high parton density corrections to
$F_2$ is rather small while they are substantial for the gluon structure
function.

\section{ 
$\mathbf{Q^2_s({x})}$ from $\mathbf{Q^2}$ - dependence of
$\mathbf{F_2}$ -
slope}
There is a hope that the $F_2$-slope ($ \frac{ \partial F_2}{
\partial \ln Q^2}$ ) will provide a measurement of the saturation scale
since in the DGLAP evolution this slope $
 \frac{\partial F_2}{\partial \ln Q^2}\,\,\,=\,\,\frac{2 \as}{9
\,\pi}\,x\,G^{DGLAP}(x,Q^2)$ directly proportional to the gluon structure
function. On the other hand, a gluon  saturation leads to the slope which
is proportional\cite{GLMSLP}  to $Q^2 R^2$ at fixed $x$  where $R$ is the
target size. Indeed, it turns out that hdQCD corrections\cite{GLMSLP} are
able to
describe all experimental data on the $F_2$-slope (see Fig.8). We consider
as an
important sign, that we are on the right track, the fact that two DGLAP
parameterizations GRV'94 and GRV'98 lead to a  good description of the
experimental data after taking into account  hdQCD effects.  

\end{itemize}
\begin{figure}
\begin{center}
\epsfig{file=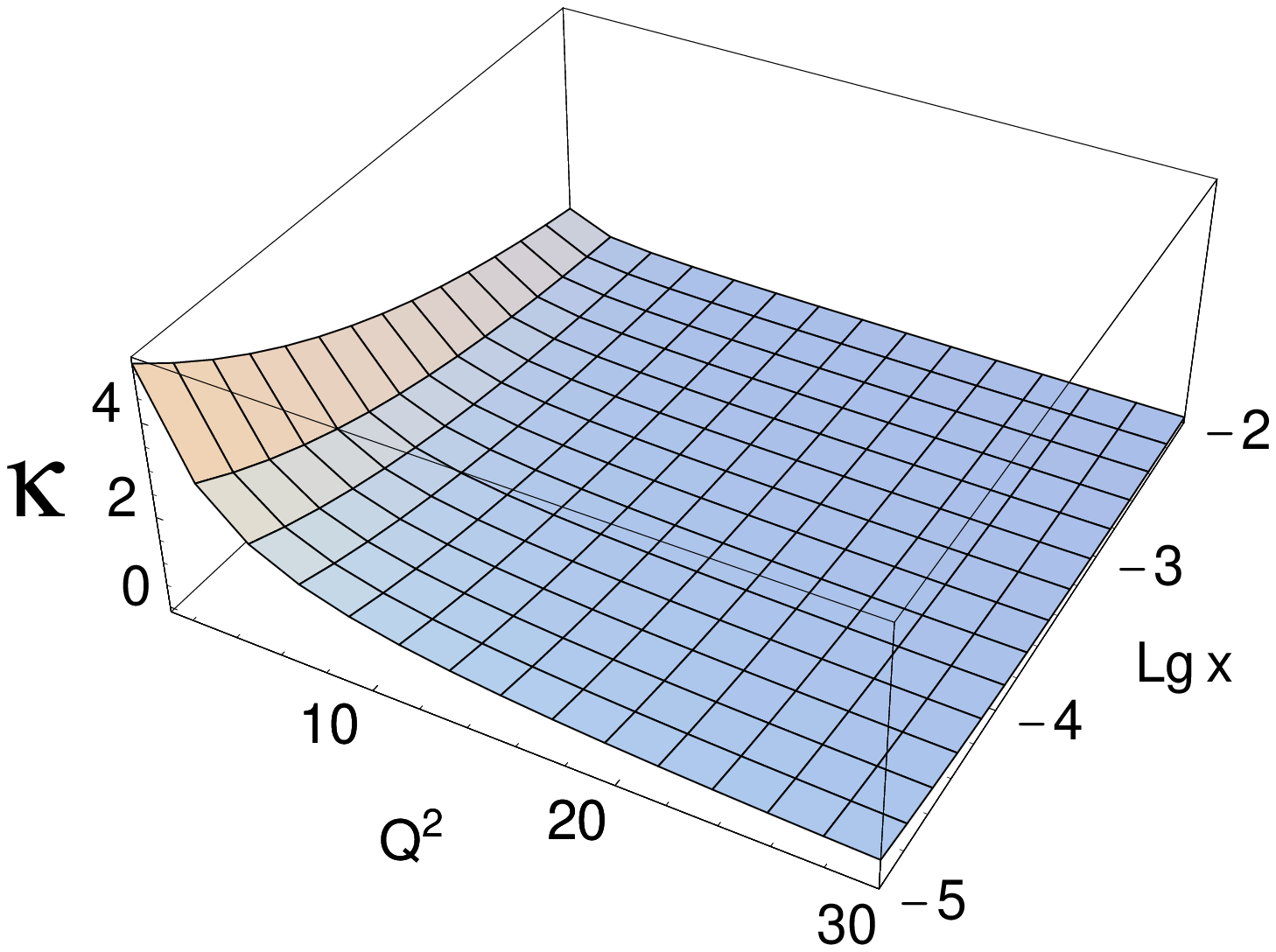, width=8cm}
\end{center}
\caption{}
\end{figure}

\begin{figure}
\vspace{-0.4cm}
\begin{center}
\epsfig{file=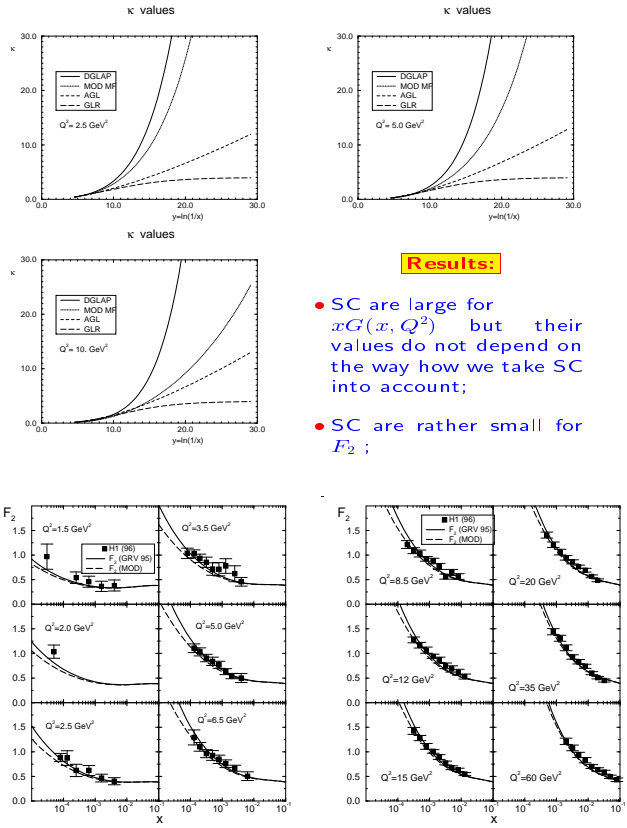,width=100mm}
\end{center}
\caption{The hdQCD corrections to $F_2$ and $xG(x,Q^2)$ calculated in Ref.
\protect\cite{AGL}. Calculation were done using \eq{EQ}.}
\end{figure}

\begin{figure}
\begin{center}
\epsfxsize=10cm
\leavevmode
\hbox{ \epsffile{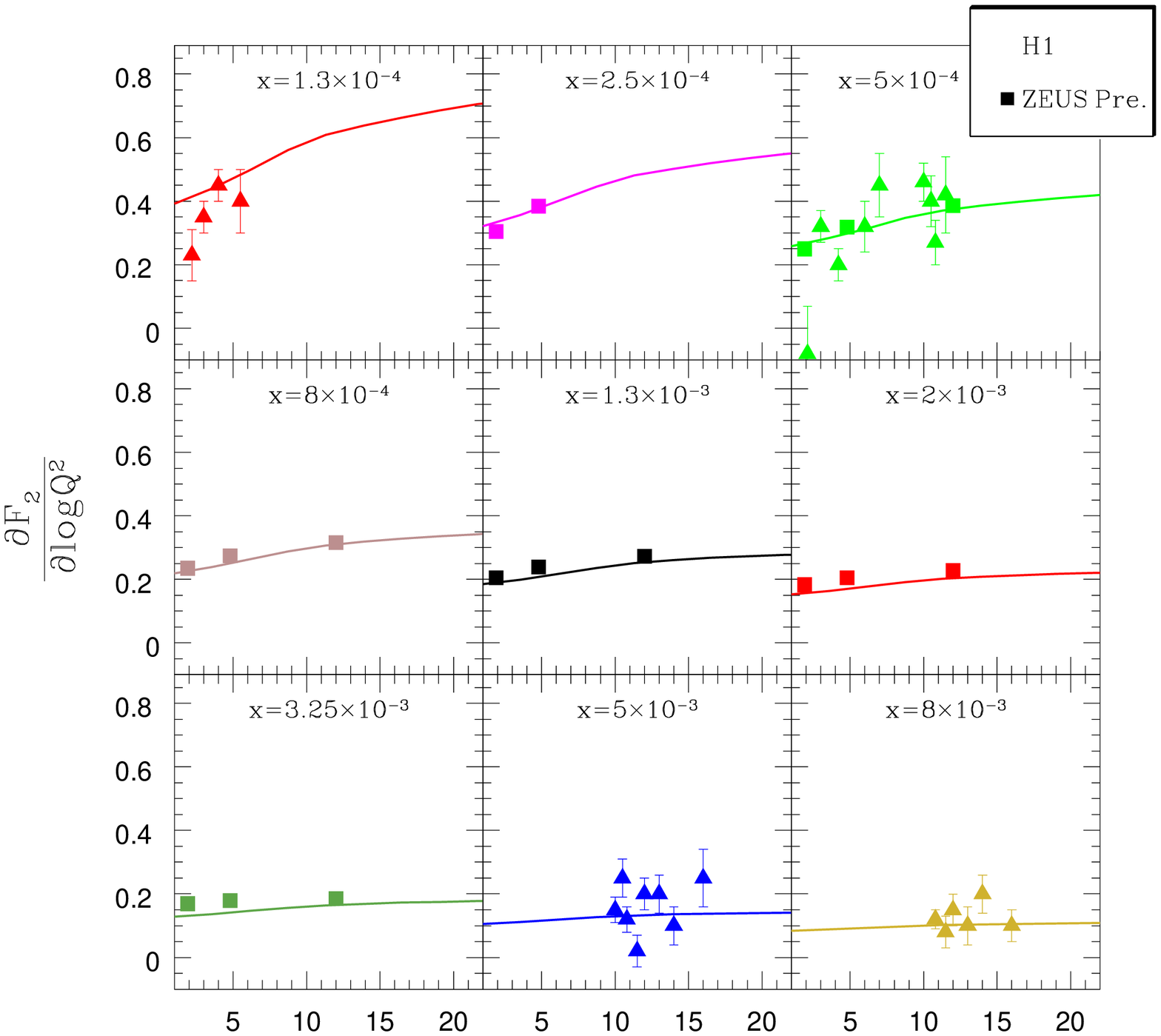}}
\end{center}
\caption{}
\end{figure}

However, the $F_2$-slope  data
 can be equally well described by ``soft" and ``hard" contribution without
hdQCD effects if we assume that the ``soft" contribution stems from rather
short distance $ 0.3 \div 0.5\,fm $ \cite{CDLONG} (see Fig.9).

\section{High density QCD and  diffractive J/$\Psi$-production in
DIS}

It turns out that it is very instructive to consider two observables : the
$F_2$- slope and the  J/$\Psi$-production in DIS.  The hdQCD effect are
essential in both observables and a simultaneous analysis of them could
give an information on the value of the saturation scale. Performing such
an analysis we obtain the following: (i)none of MRS parameterizations
survives; (ii)all GRV parameterizations  survive only with hdQCD effects;
and (ii) $\chi^2$/n.d.f. is excellent for GRV + hdQCD effects (see Fig. 10
).

\begin{figure}
\begin{center}
\epsfxsize=10cm
\leavevmode
\hbox{ \epsffile{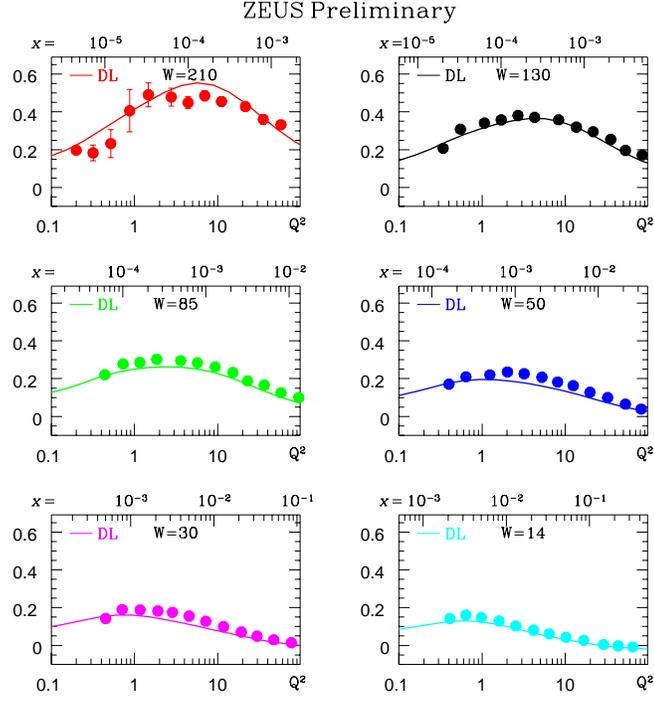}}
\end{center}
\caption{ The $F_2$ -slope in D-L model with ``soft" contribution at
rather short distances \protect\cite{DL2P}}
\end{figure}

\begin{figure}   
\begin{center}
 \epsfig{file=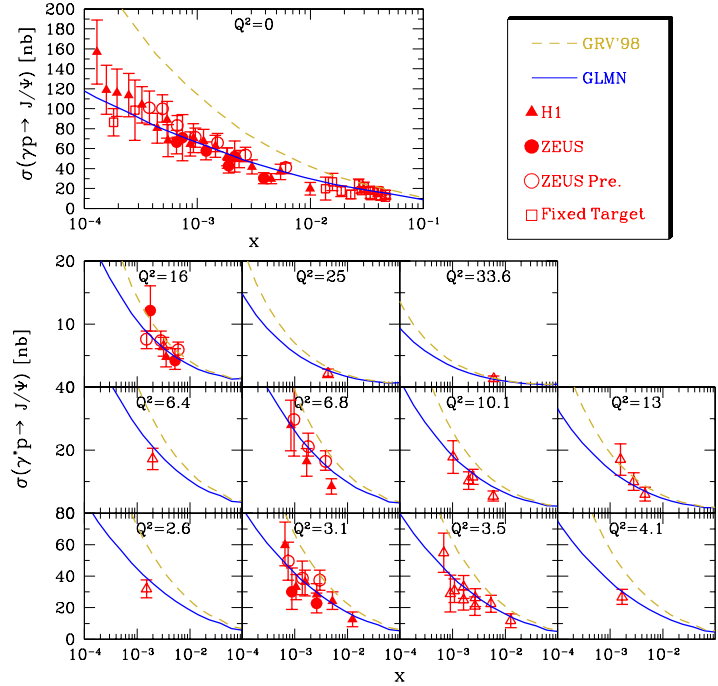,width=100mm}
\end{center} 
\caption{J/$\Psi$ production in hdQCD approach. Curves are taken from
Ref. \protect\cite{OSAKA}}
\end{figure}

\section{A new scaling }  

The saturation hypothesis got a new support by Golec-Biernat and
W\"{u}sthoff \cite{GW} who suggested a simple model that introduces only
one scale
for colour dipole - target interaction: the saturation scale $Q_s(x)$ ,
which they defined by fitting experimental data.
This simple model does not take into account even such well established
property as scaling violation but describes all experimental data from
HERA for $x < 0.01$. 

 In Ref. \cite{SKG} it was noticed that the HERA data 
show a new scaling: $\sigma( \gamma^*  p ) = R^2 F(Q^2/Q^2_s(x))$ for all
values of $Q^2$ at $x < 0.01$ ( see Fig. 11 ). Such a scaling was expected
\cite{GLR,BL,LV,LK} in the saturation region to the left of the critical
line
$\kappa=1$ in Fig.2. The value of the saturation scale $Q_s(x)$ can be
measured as a value of $Q^2$ at which we see  a deviation from this
scaling. The fact, that all data at HERA kinematic region for $x < 0.01$
show this scaling,  confirms the idea that at $x < 0.01$ DIS is deeply in
the saturation region. It should be stressed that the whole idea of
saturation is the simple fact that the only one scale $r_{saturation}$
determines the scattering amplitude for the distances longer than
 $r_{saturation}$. This very fact one can see directly in Fig.2,
noticing that a hadron looks as the diffraction grid with the size
$r_{saturation}$ in the saturation region ($\kappa \gg 1 $ in Fig. 2 ).

\begin{figure}
\begin{center}
\epsfxsize=13cm
\leavevmode
\hbox{ \epsffile{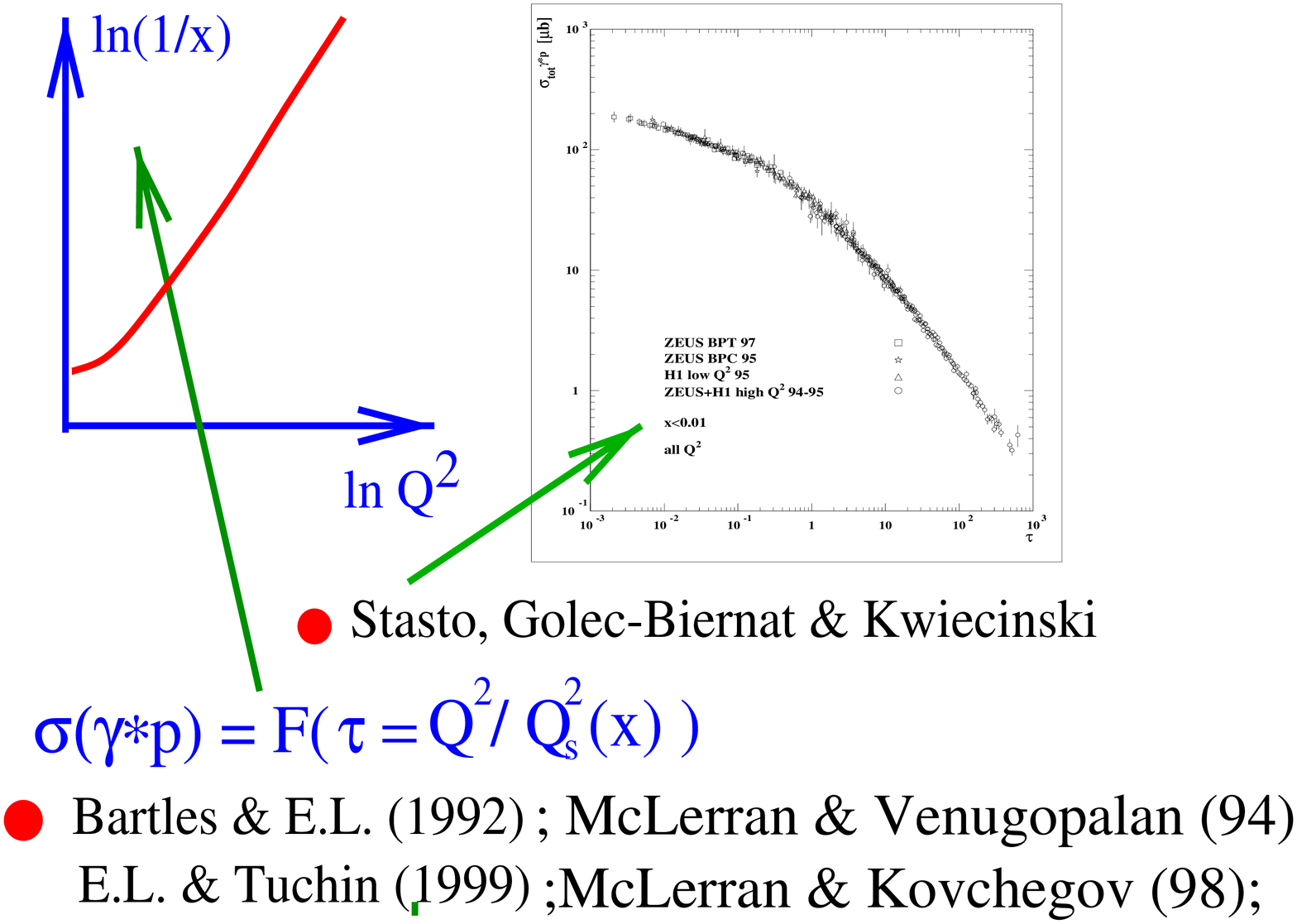}}
\end{center}
\caption{}
\end{figure}

\section{Summary}     
 
The brief review on low $x$ physics, which is given in this talk, allows
us to make a definite conclusion, that a new QCD regime has been reached
at HERA: the regime of high parton density QCD with a gluon  saturation. 
Let us list all arguments for  such  a  new regime:
\begin{itemize}
\item \quad  There are no experimental data that are in a contradiction
with the asymptotic prediction of high density QCD. Indeed,
\begin{itemize}
\item \quad The  $F_2$-slope shows $ d F_2/d ln Q^2 \propto Q^2 $
behaviour
at $Q^2 < Q^2_s(x)$ \cite{ZEUSSLP,H1SLP};
\item \quad The ratio of diffraction cross section in DIS to the total DIS
cross section is constant at HERA kinematic region \cite{LK};
\item \quad The inclusive diffraction stems from short distances as should
be in a gluon saturation picture in which a hadron looks as a diffraction
grid with a typical size $ 1/Q_s(x)$ \cite{ZEUSDD,H1DD};
\end{itemize}
\item \quad The HERA data show a new scaling\cite{SKG}, predicted
theoretically \cite{GLR,BL,LV,LK} in the saturation region , namely, that 
$\sigma( \gamma^*  p ) = R^2 F(Q^2/Q^2_s(x))$ for all
values of $Q^2$ at $x < 0.01$. 
\item \quad All data can be described in the simplest saturation model of 
Golec-Biernat and W\"{u}sthoff \cite{GW};
\item \quad The current parameterizations, based on  the DGLAP evolution
equations, cannot describe simultaneously
the $Q^2$-behaviour of the $F_2$ slope and energy behaviour of the
J/$\Psi$ production in DIS and photoproduction\cite{OSAKA};
\item \quad The theoretical developments in high density QCD has been so
remarkable during the past two years, that we can trust \eq{EQ} which
predicts the essential high density collective phenomena in HERA
kinematic region;
\item \quad  The simple estimates based on the gluon density extracted
from the HERA data show large packing factor (see Fig. 6 ) or, in other
words, they show the strong shadowing corrections are needed in HERA
kinematic region.
\end{itemize}

In spite of everything mentioned above, not everybody will agree with the
strong statement that has been formulated here. The reason for this is
very simple: most of data, that has been considered above, have  different
explanations without SC. All these alternative explanations cannot be
considered as natural ones, but it is behind many people scepticism on a
new QCD regime at HERA, that the DGLAP evolution has more fundamental
origin in QCD than all estimates with SC. It is not true at all and SC are
more general approach than the DGLAP equation because (i) they are
consistent with the $s$-channel unitarity; and (ii) the non-linear
equation (see \eq{EQ} )has the same
deep operator proof as the DGLAP evolution equation.

{\bf Acknowledgments:}

I wish to thank  E. Gotsman, M. Lublinsky, U. Maor, E. Naftali and K.
Tuchin for a pleasure to work with them and to discuss with them all
problems  of low $x$ physics.

This research was supported by part by Israel Academy of Science and
Humanities and by BSF grant \#
98000276.


\begin{thebibliography}{99}
\bibitem{GLR}
L.V. Gribov, E.M.  Levin and M.G.  Ryskin, {\it Phys. Rep}
{\bf 100} (1983) 1.
\bibitem{MUQI}
A.H. Mueller and J. Qiu, {\it Nucl. Phys.} {\bf B268} (1986) 427.
\bibitem{LV}
L. McLerran and R. Venugopalan,{\it Phys. Rev. } {\bf D49} (1994)
2233,3352, {\bf 50} (1994) 2225, {\bf 53} (1996) 458, {\bf 59} (1999)
094002.  
\bibitem{GRIB}
V.N. Gribov, {\it Sov. Phys. JETP} {\bf 30} (1970) 709.
\bibitem{MU94}
A.H.  Mueller, {\it  Nucl. Phys.}  {\bf B415} (1994) 373.
\bibitem{AQM}
E. Levin and L. Frankfurt, {\it JETP Lett.} {\bf 2} (1965) 65;
H. J. Lipkin and F. Scheck, {\it Phys. Rev. Lett.} {\bf 16 } (1966) 71.
\bibitem{CDLONG}
J. R. Forshaw, G. Kerley and G. Shaw, {\it Phys.Rev.} {\bf D60} (1999)
074012, {\it Nucl. Phys.} {\bf A675} (2000) 80;\\
M. McDermott, L. Frankfurt, V. Guzey and M. Strikman,{\it "Unitarity and
the QCD-improved dipole picture"}, {\tt hep-ph/9912547}.
\bibitem{PQCD}
E.   Levin and M.G. Ryskin, \prep{189}{267}{1990};\\
J.C.Collins and J. Kwiecinski, \npb{335}{90}{89};\\
J. Bartels, J. Blumlein and G. Shuler, \zpc{50}{91}{91};\\
E. Laenen and E. Levin, \arnps{44}{94}{199}
and references therein;\\
A.L. Ayala, M.B. Gay Ducati and E.M. Levin, \npb{493}{97}{305},
{\bf B510} (1990) 355;\\
Ia. Balitsky, {\it Nucl.Phys. } {\bf B463}  (1996) 99;\\
Yu. Kovchegov, \prd{54}{1996}{5463}, {\bf D55}(1997) 5445,
{\bf D60}(2000) 034008,
{\bf D61} (2000)074018;\\ A.H. Mueller,
{\it Nucl. Phys.} {\bf B572} (2000) 227,
{\bf B558} (1999) 285;\\ Yu. V. Kovchegov, A.H. Mueller,
\npb{529}{98}{451};\\
 E.  Levin
and
K. Tuchin, {\it Nucl. Phys.} {\bf B573}(2000) 833.

\bibitem{NPQCD}
J. Jalilian-Marian, A. Kovner, L. McLerran  and  H.
Weigert, \prd{D55}{97}{5414};\\
J. Jalilian-Marian, A. Kovner and  H.
Weigert, \prd{59}{99}{014015};\\
J. Jalilian-Marian, A. Kovner and  H.
Weigert, \prd{59}{99}{014015};\\
J. Jalilian-Marian, A. Kovner, A.
Leonidov and  H. Weigert, \prd{59}{99}{014014,034007},
Erratum-ibid. \prd{59}{99}{099903};\\ 
A. Kovner, J.Guilherme Milhano and  H. Weigert,
OUTP-00-10P,NORDITA-2000-14-HE, {\tt hep-ph/0004014};\\
 H. Weigert, NORDITA-2000-34-HE, {\tt hep-ph/0004044}.
\bibitem{EQ}
Ia. Balitsky, {\it Nucl.Phys. } {\bf B463}  (1996) 99;
Yu. Kovchegov,
\prd{60}{1000}{034008}.
\bibitem{SEPSC}
  E. Gotsman, E. Levin and  U. Maor,  { \it Eur.Phys.J.}
{\bf C5} (1998) 303;
A.D. Martin, M.G. Ryskin and  A.M. Stasto, {\it
Eur.Phys.J.} {\bf C7} (1999) 643;
 E. Gotsman, E. Levin, U. Maor and  E. Naftali, {\it
Eur.Phys.J.} {\bf C10} (1999) 689.
\bibitem{KB}
 B. Badelek and J. Kwiecinski, {\it Phys.Lett.} {\bf B295}
(1992) 263.
\bibitem{HERAREV}
A. M. Cooper-Sarkar, R. C. E. Devenish and A. De Roeck, {\it Int. J. Mod.
Phys.} {\bf A13} (1998) 33;\\
 H. Abramowicz and A. Caldwell, {\it Rev.
Mod.
Phys.} {\bf 71} (1999) 1275.
\bibitem{GLMSLP}
E. Gotsman, E. Levin, U. Maor and  E. Naftali,
{\it Nucl.Phys.} {\bf B539} (1999) 535;
 E. Gotsman, E. Levin and  U. Maor, {\it  Phys.Lett.}
{\bf B425} (1998) 369.
\bibitem{AGL}
A.L. Ayala, M.B. Gay Ducati and E.M. Levin, \npb{493}{97}{305},
{\bf B510} (1990) 355.

\bibitem{DL2P}
A. Donnachie and P.V. Landshoff:\plb{437}{98}{408}; \plb{470}{99}{243}.   

\bibitem{OSAKA}
E. Gotsman, E. Ferreira, E. Levin, U. Maor and  E. Naftali, {\it ``
Screening corrections in DIS at low $Q^2$ and $x$" },
Talk given at 30th International Conference on High-Energy Physics
(ICHEP 2000), Osaka, Japan, 27 Jul - 2 Aug 2000, {\tt  hep-ph/0007274}.
\bibitem{GW}
K. Golec-Biernat and M. Wusthoff, {\it Phys. Rev.} {\bf D59} (1999) 
014017; {\bf D60} (1999) 114023;
K.Golec-Biernat,{Talk
at 8th International Workshop on Deep Inelastic Scattering and
QCD (DIS 2000)}, Liverpool, England, 25-30 Apr 2000,{\tt  hep-ph/0006080}.
\bibitem{SKG}
A.M. Stasto, K. Golec-Biernat and  J. Kwiecinski, {\it `` Geometric
scaling for the total $\gamma^* p$ cross-section in the low $x$ region"},
{\tt hep-ph/0007192}.
\bibitem{BL}
J. Bartels and E. Levin,{\it  Nucl.Phys.} {\bf B387} (1992) 617.
\bibitem{LK}
Yu.V. Kovchegov and L. McLerran,  { \it Phys.Rev.} {\bf D60}
(1999) 054025.
\bibitem{ZEUSSLP}
B. Foster (ZEUS Collaboration),  Invited talk, Royal Society Meeting,
London, May 2000;\\
ZEUS collaboration: J. Breitweg et al., {\it Phys. Lett.} {\bf B487}
(2000) 53.
\bibitem{H1SLP}
M. Klein (H1 collaboration), {\it
``Structure Functions in Deep Inelastic Lepton-Nucleon Scattering"}
   Talk at Lepton-Photon  Symposium, Stanford, August 1999, {\tt
hep-ex/0001059}.
\bibitem{ZEUSDD}
ZEUS Collaboration: J.  Breitweq et al., {\it Eur. Phys. J.} {\bf C6}
(1999) 43.
\bibitem{H1DD}
H1 Collaboration: C. Adloff et al.,  \zpc{76}{97}{613}.

\end{thebibliography}
\end{document}